# Atomic Clouds as Spectrally-Selective and Tunable Delay Lines for Single Photons from Quantum Dots

*Johannes S. Wildmann*[†,]*, *Rinaldo Trotta*[†]**, *Javier Martín-Sánchez*[†], *Eugenio Zallo*[‡], *Mark O' Steen*[§], *Oliver G. Schmidt*[‡], *and Armando Rastelli*[†]

[†]Institute of Semiconductor and Solid State Physics, Johannes Kepler University, Altenbergerstr. 69, A-4040 Linz, Austria

[‡]Institute for Integrative Nanosciences, IFW Dresden, Helmholtzstr. 20, D-01069 Dresden, Germany

[§]Veeco Instruments, 4875 Constellation Drive, 55127 St. Paul, Minnesota, USA

We demonstrate a compact, spectrally-selective, and tunable delay line for single photons emitted by quantum dots. This is achieved by fine-tuning the wavelength of the optical transitions of such "artificial atoms" into a spectral window in which a cloud of natural atoms behaves as slow-light medium. By employing the ground-state fine-structure-split exciton confined in an InGaAs/GaAs quantum dot as a source of single photons at different frequencies and the hyperfine-structure-split D1 transition of Cs-vapors as a tunable delay-medium, we achieve a differential delay of up 2.4 ns on a 7.5 cm long path for photons that are only 60 μeV (14.5 GHz) apart. To quantitatively explain the experimental data we develop a theoretical model that accounts for both the inhomogeneously broadening of the quantum-dot emission lines and the Doppler-broadening of the atomic lines. The concept we proposed here may be used to implement time-reordering operations aimed at erasing the "which-path" information that deteriorates entangled-photon emission from excitons with finite fine-structure-splitting.

Optically active epitaxial quantum dots (QDs) have emerged as efficient sources of single [1] and entangled photons [2,3] on demand [4] with potential applications in the field of quantum communication[5]. By interfacing photons emitted by QDs with clouds of natural atoms, photon storage at single photon level[6] may become possible, thus opening the route to the realization of hybrid interconnects for quantum networking[7]. Pioneering experiments in this field [8] have shown slow-light in Rb clouds using single photons emitted by GaAs QDs. These experiments – based on the original concept proposed by Camacho et al. [9,10] – were performed by tuning the QD emission lines between the hyperfine-split $D_2$ transition of $^{87}$Rb. Despite the work clearly shows the potential of the hybrid natural-artificial interface, the pronounced spectral broadening of the QD-photon source employed for the experiments prevented a detailed analysis of the temporal delay as a function of the relative spectral position of QD and atomic transitions. This is an interesting aspect of the hybrid interface because it could allow introducing temporal delays between photons whose frequencies differ only by a few GHz. The resulting spectrally-selective delay-line could be exploited not only as a filter in experiments aiming at storing and retrieving single photons [6], but it could represent a useful tool to "reorder" in time the temporal sequence of photons originating from radiative cascades in real [11] and artificial atoms [12,13].

In this work, we demonstrate that a cloud of cesium atoms can be used to introduce a significant temporal delay (up to 2.4 ns for transitions featuring lifetimes $\tau_{QD} \approx 1ns$) between photons which are separated in frequency by more than 10 GHz. As source of single photons with different colors we use the fine-structure-split emission lines of excitons confined in single self-assembled InGaAs/GaAs QDs. Their energy can be finely adjusted to the middle of the $D_1$ transitions of Cs-vapors via external electric or strain fields provided by diode-like nanomembranes[14,15] integrated onto piezoelectric actuators [16,17]. We show that the amount temporal delay[18] between the photons can be tuned by varying the temperature of the Cs cell and that the “antibunched” character of the quantum source is retained after photon-propagation though the cell. Finally, we develop a theoretical model that quantitatively explains all the experimentally observed features of the spectrally-selective delay line.

Figure 1a illustrates the idea behind the experiment performed in this work. We use a single QD as a source of photons with different frequency (see the blue and the red photons in the Figure 1a) and we employ external strain or electric field to tune their emission energy across the hyperfine-split $D_1$ lines of a Cs clouds contained in a quartz cell. The photon interaction with Cs atoms in the cell (i.e., the time delay) depends eventually on their energy. When the photon energy matches the center of the $D_1$ lines, the photon group velocity can be drastically decreased with respect to the vacuum value. Therefore, the time sequence of photons emitted by QDs with slightly different frequencies can be controlled by simply varying the temperature of the cell, which increases the optical depth of the absorption line and, due to a change in the real part of the refractive index, modifies also the group velocity.

We start out characterizing the properties of the quartz cell containing the cesium cloud. The Cs D1 transitions are split into the $6^2S_{1/2}$ and the $6^2P_{1/2}$ levels, which are further split due to the hyperfine coupling into levels characterized by total atomic angular momentum F=3 and F=4. Therefore, there are 4 possible transitions highlighted in Figure 1d with different colored arrows. [19]. The optical transmission measurements around the D1 lines through the Cs cell are reported in Figure 1b for different vapor temperatures (for details on the measurements see the supplementary information). The characteristic four transitions of the hyperfine structure are clearly resolved for a cell temperature $T_{cell}$=70 °C (black line), but they quickly broaden as $T_{cell}$ is increased due to well-known Doppler broadening [20]. Most importantly, two absorption dips separated by 10 GHz remain for $T_{cell}$>100 °C. This splitting is considerably larger than the one of the D2 lines of Rb[8] $\approx 6.8\ GHZ$ and is therefore more suitable for experiments involving spectrally-broadened QD lines (see the following).

Based on these measurements we are able to simulate the optical response of our Cs cell in the proximity of the $D_1$ lines. The numerical simulations make use of the susceptibility [9, 18] and take into account all possible transitions of the Cs $D_1$ line. More specifically, the susceptibility of a medium with 4 resonances ($\nu_{33}$,$\nu_{34}$,$\nu_{43}$,$\nu_{44}$) can be modelled as:

$$\chi(\nu)=A\left(\frac{g_{33}}{\nu_{33}-\nu-i\,\gamma}+\frac{g_{34}}{\nu_{34}-\nu-i\,\gamma}+\frac{g_{43}}{\nu_{43}-\nu-i\,\gamma}+\frac{g_{44}}{\nu_{44}-\nu-i\,\gamma}\right)$$

where $g_{33}$,$g_{34}$,$g_{43}$,$g_{44}$ are the oscillator strengths of each resonance [19] and a damping constant $\gamma$, which determines the linewidth of the resonances. Finally, the complex refractive index n and group velocity $v_g$ can be derived from the susceptibility giving:

$$v_g=\frac{1}{n(\nu)+\nu\frac{d\,n(\nu)}{d\,\nu}}$$

The simulated $v_g$ is shown in Figure 1c for different temperatures as a function of the detuning ($\Delta$) with respect to the Cs D1transition. For small $\Delta$ we find that $v_g$ is significantly reduced with respect to the speed of light (c) while there is practically no absorption of the photons propagating through the cell (see Figure 1b). It is also interesting to note that it is not necessary to tune the energy of the photons exactly in the middle of the hyperfine doublet to observe slow-light, as can be clearly seen in Fig. 1b for $\Delta$ larger than 5 GHz. Finally, we note that the magnitude of the delay can be easily increased by changing temperature of the cell, revealing that $v_g$ can become negative at the absorption lines. This interesting effect originates from an anomalous dispersion and can either lead to fast light or a backward propagating wave[21].

Having characterized our slow-light medium we now address the transmission through the Cs cell of streams of photons emitted by semiconductor QDs. We use InGaAs QDs embedded in strain-tunable optoelectronic devices, where large strain and electric fields are used to fine tune the energy of optical transitions across the spectral region of the $D_1$ line of Cs. Details on sample fabrication and device performances can be found elsewhere[22].

Figure 2a shows color-coded micro-photoluminescence spectra of a negatively charged exciton (trion) that is tuned through the hyperfine structure of Cs D1 by varying the electric field $F_p$ across the piezo (i.e. the QD strain status, see Ref 16,17 for details). A strong quenching of the transmitted light is observed for $F_p = 7.05$ and $7.5$ kV/cm as result of optical absorption in the Cs vapor. This effect can be better observed in the Figure 2b, where the intensity of the QD light transmitted through the Cs cell is reported as a function of $F_p$ and $\Delta$ (calculated assuming a linear relation between the emission energy shift and $F_p$, see Ref. 16). Considering that the temperature of the Cs cell is 135 °C, the data of Fig. 2b nicely match those of Fig.1b but for the additional broadening of the transmission dips. This is due to the broad linewidth of the trion transition, which could not be resolved using our spectrometer (featuring a spectral resolution of 25 μeV). By performing a convolution between the measured Cs transmission spectrum at 136 °C (see Fig. 1b) and a Gaussian function we find the best agreement to the experimental data via least squares minimization and using the linewidth ω as the only simulation parameter (see the supplemental material). Using this procedure, we find $\omega = 10.3 \pm 0.1$ μeV. On the one hand, this analysis clearly shows that sweeping the QD lines through the hyperfine structure of atomic clouds is a useful tool for high-resolution spectroscopy. On the other hand, the value of ω we measure highlights the improved optical quality of our QDs compared to those used in Ref 23, meaning that QD line can be conveniently tuned to the middle of the D1 transitions without substantial photon absorption. This allows us to demonstrate that the antibunching character of the photon source is retained under insertion of a slow-light medium in the optical path. Fig. 2c shows autocorrelation measurements for photons emitted by a different QD. For these measurements we used a trion featuring $\tau_{QD}$=1.04$\pm$0.1 ns and $\omega_{QD}$=26.1 $\pm$ 0.5 μeV. The red (black) curve in Figure 2c shows the result of the experiment for $\Delta > 15$ GHz , ($\Delta = 0$). Ideally, the autocorrelation histogram for a perfect single photon source should display a series of peaks of equal amplitude separated by the inverse of the excitation laser frequency (80 MHz here) and a missing peak at zero time-delay. On the one hand, the measurements for detuning $\Delta > 15$ GHz shows clearly photon-antibunching and that the source is a single quantum emitter. On the other hand, it also shows a multi-photon emission probability of $g^{(2)}(0)= 0.17\pm0.08$. This value most probably arises from carrier recapture phenomena on a time scale comparable with the trion lifetime[24]. The autocorrelation histogram for $\Delta$=0 shows (*i*) a very similar multiphoton emission probability of $0.21\pm0.12$, (*ii*) a reduced number of coincidence counts and (*iii*) a broadening of the autocorrelation peaks. While (*i*) clearly indicates that the antibunched-nature of the single photon source is unaffected by the insertion of Cesium vapor in the optical path, (*ii*) and (*iii*) can be explained by the broadening of the trion transition (26.1 μeV), which results in higher photon absorption in the Cs vapor and in a different shape in time of the photon wavepacket [10], as discussed in more details below.

We now demonstrate the possibility of using a Cs vapor cell as a spectrally-selective delay line. For these measurements we use two fine-structure-split emission lines of a neutral exciton from a different QD and we tune them in energy via the quantum-confined Stark resulting from changing the electric field $F_d$ across the p-i-n diode structure (see Ref. 25 for more details). The vertical (V) and horizontal (H) polarized PL spectra of the neutral exciton are shown in Figure 3a, which reveals a fine structure splitting of 59 μeV, a value larger than the splitting between the two absorption lines of Cs $D_1$ (41 μeV, see Figure 1b). The PL intensity for each orthogonally-polarized component is shown in Figure 3b as a function of $F_d$. The two absorption maxima related with the Cs D1 hyperfine structure can be clearly resolved for both components and – as expected – happen to be at different $F_d$. This means that, in principle, the temporal sequence of the differently-polarized photons escaping from the Cs cell can be varied according to their energetic position with respect to the $D_1$ transitions of the Cs vapor. Fig. 3c-f shows time-resolved PL measurements at different Cs vapor cell temperatures for $F_d$ = -58.7 kV/cm (H-polarized photons tuned to Cs D1, light blue) and $F_d$ = -57.6 kV/cm (V-polarized photons tuned to Cs D1, light red). There are two main effects: *i*) both H and V-polarized photons can be independently delayed with respect to the detuned case; *ii*) the time delay strongly depends on the cell temperature, being larger at higher $T_{cell}$. More specifically, we measure a shift from 3.2 ns ($T_{cell}$=101.5 °C) to 5.1 ns ($T_{cell}$=143 °C) in the case of H-polarized tuned photons, and from 2.9 ns ($T_{cell}$=101.5 °C) to 4.9 ns ($T_{cell}$=143 °C) in the case of V-polarized tuned photons. Negligible time delays were instead measured for all cell temperatures when both H-polarized and V-polarized photons are detuned by $\Delta$ = 14.3 GHz ≈59 μeV. Besides the temporal delay, we observe a clear change in the temporal distribution of the transmitted photons, as can be seen by comparing the exponential decays of Figure 3c and 3d. This finding can be qualitatively explained considering the combined effects of the inhomogeneously broadened QD emission ( ω=6.5 GHz) and the Doppler broadened absorption of Cs that produce a dispersion of the group velocity and the transmission through the vapor. In order to quantitatively account for the experimental data, we model the QD emission in frequency with a Gaussian distribution featuring the spectral linewidth extracted from Fig. 3b while, for the time-resolved measurement, we use an exponentially modified Gaussian distribution [26] (as see supporting information). The temporal distribution of the photon wavepacket after propagation through the Cs cell at a specific temperature can be then calculated by discretizing the spectrally-broadened QD lines and evaluating (for different detuning $\Delta$) the expected delay and relative intensity. Finally, by integrating over all $\Delta$, the time-traces as a function of the cell temperature, QD linewidth and detuning can be obtained. Figure 4shows the result of the simulations for $\Delta=0$ (Figure 4a) and 59 μeV ($\Delta=14.3$ GHz) detuning (Figure 4b). In both cases, we are clearly able to reproduce all the features of the experimental data, thus confirming that the reshaping in time of the photon wavepacket arises from the convolution of the spectrally-broadened QD emission with the different group velocities and rate of absorption in Cs vapors. For a better comparison between simulated and experimental results, we extracted the values of lifetime and delays see, respectively Figure 4d and 4c. The $\Delta = 0$, the delay of the two excitonic lines (green and blue lines) follow an exponential function with temperature, a trend which is well reproduced by our simulations (see the dashed line).We are also able to reproduce the small and apparently linear increase of the delay with temperature for the detuned case (pink and light blue). In addition to the shift in time, we investigated the broadening of the photon-wavepacket by considering the changes in the decay time, i.e., the lifetime, see Figure. 4d. The data extracted from the experiments, presented as symbols connected by solid lines, show that for $\Delta = 0$ (blue, green), the lifetime tends to increase with increasing cell temperature, while it is almost unaffected for $\Delta$=14.3 GHz (pink, light blue). Considering the scatter in the experimental data, the observed trends are reproduced fairly well by the simulations (dashed lines).

In summary, we have demonstrated that it is possible to use warm Cs vapors as a spectrally-selective and tunable delay lines for single photons emitted by semiconductor quantum dots. By performing autocorrelation measurements we show for the first time that the anti-bunched character of the QD photon source is retained after photon propagation through a slow-light medium made out of an atomic vapor. Then, we show that it is possible to introduce a significant temporal delay between photons that are spectrally separated by only few GHz. By tuning via external strain or electric fields the energy of the two fine-structure-split excitons through the hyperfine structure of the D1 line of Cs, we were able to slow down independently photons originating from each transition. The measured temporal delay and time-distribution of the photon wavepackets are quantitatively explained by a model that accounts for both the inhomogeneously broadened QD emission and the Doppler-broadening of the atomic lines. The spectrally-selective delay line we present in this work can be used to reorder the arrival time of photons emitted during the biexciton-exciton radiative cascade. When used with a QD with suppressed biexciton binding energy [25] and featuring a large fine structure splitting, it may be used to experimentally demonstrate the feasibility of the recently proposed and not yet experimentally demonstrated "time-reordering" scheme for entangled photon generation.

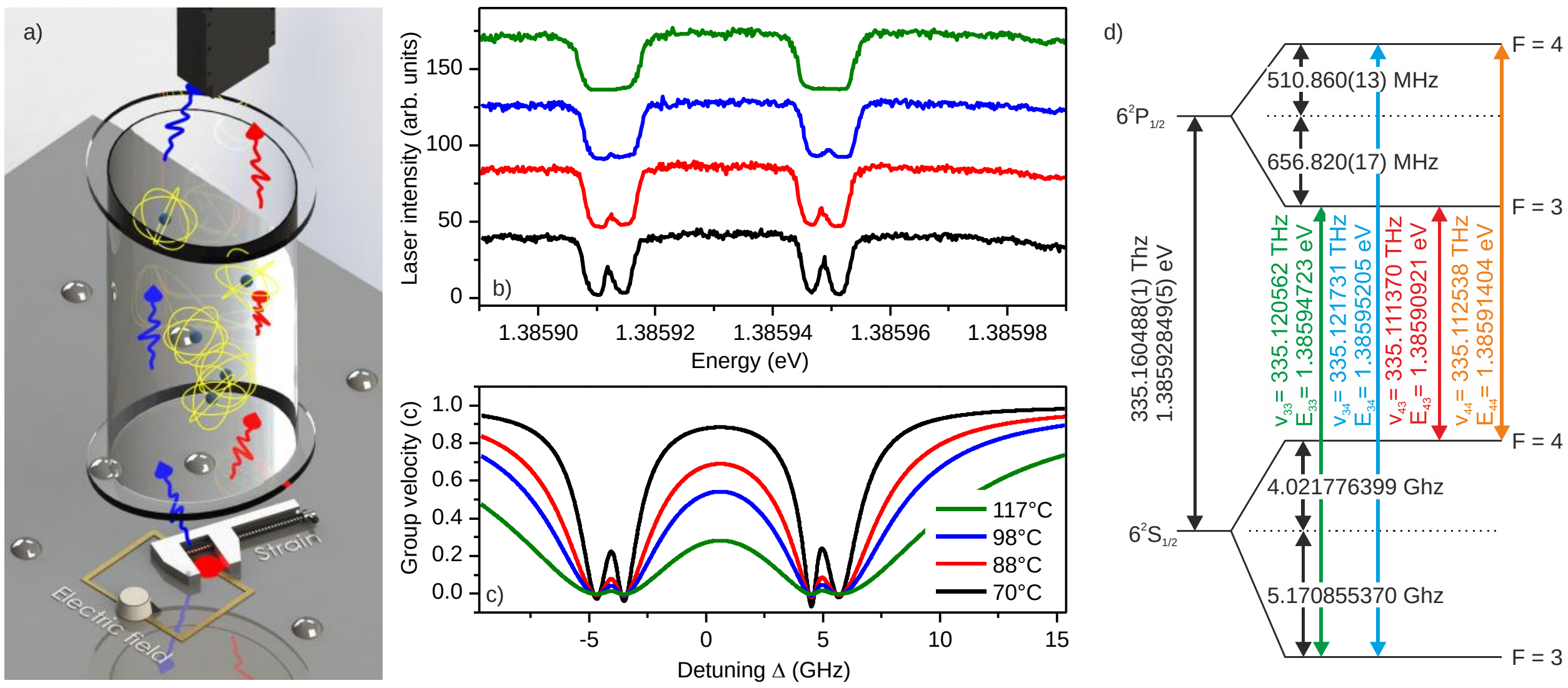

**Figure 1.** (a) Sketch of an InGaAs QD emitting single photons whose energy can be tuned by external stress or electric fields in the spectral region where a Cs vapor acts as slow-light medium. Among the photons that pass through atomic vapor, only those with specific energy (the red photons) are delayed in time. The arrival time of photons at different energies can be detected using a single photon avalanche photodiode, shown in black in the figure. (b) Transmission spectrum of Cs $D_1$ absorption lines as a function of the cell temperature. (c) Simulation of group velocity in the proximity of Cs $D_1$ for different temperatures assuming susceptibility with 4 resonances. (d) Sketch of the allowed transitions of Cs $D_1$ (indicated by vertical arrows) including hyperfine splitting in the Cs fine structure between the $6^2S_{1/2}$ and the $6^2P_{1/2}$ levels.

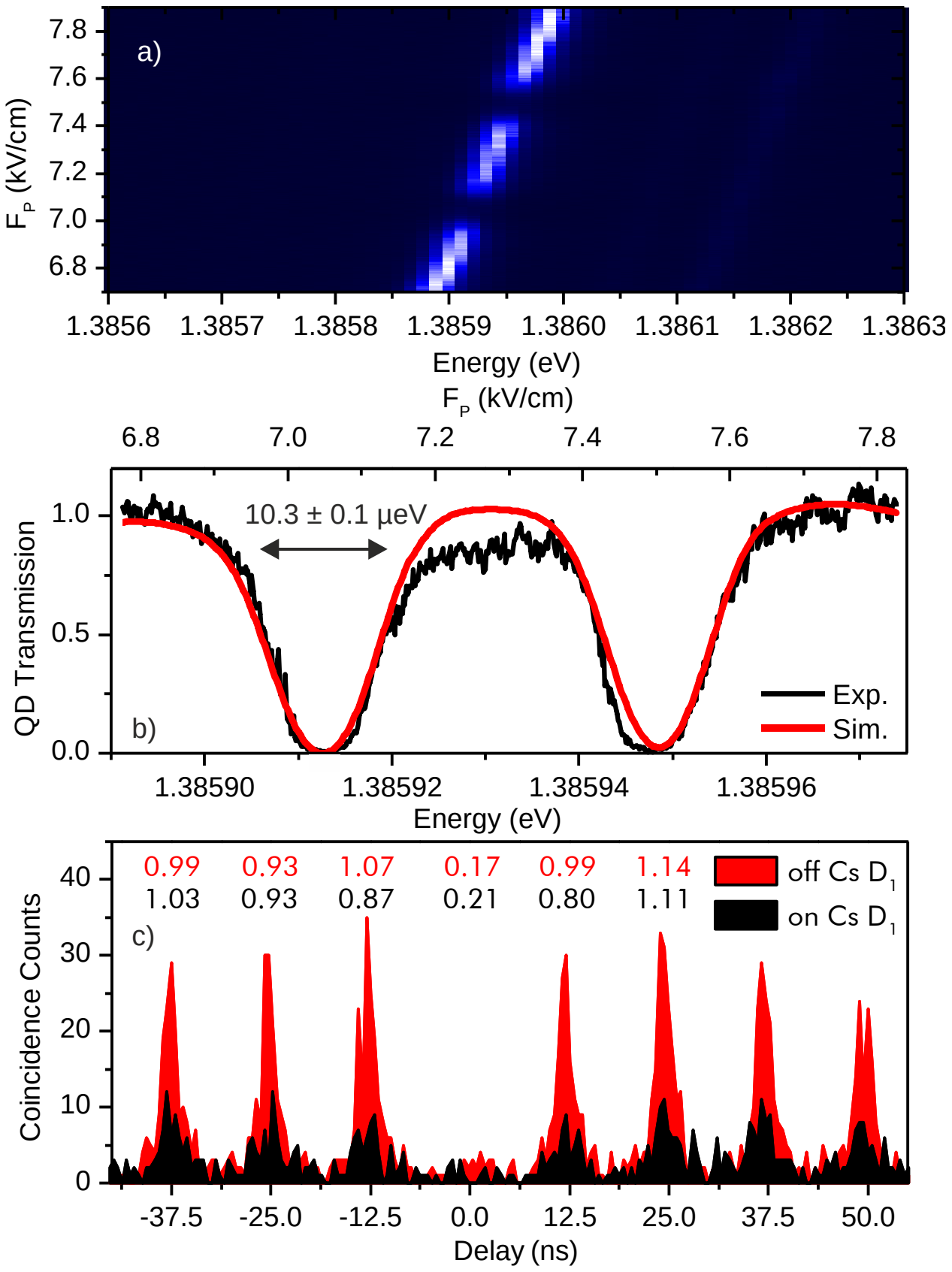

**Figure 2.** (a) Color-coded photoluminescence (PL) spectra of a negatively charged exciton (trion) which is scanned through the Cs $D_1$ absorption by the application of a variable stress generated by an electric field $F_p$ applied to piezoelectric actuator. PL transmission is strongly suppressed at 7.1 and 7.5 kV/cm. (b) Amplitude of the PL as function of $F_p$ revealing two transmission dips. The linewidth of the trion line can be obtained by deconvolving these dips with the small Hyperfine splitting of Cs $D_1$, as explained in the text. (c) Auto correlation measurement of the trion transition for $\Delta > 15\ GHz$ (red curve) and $\Delta = 0$ GHz (balck curve).

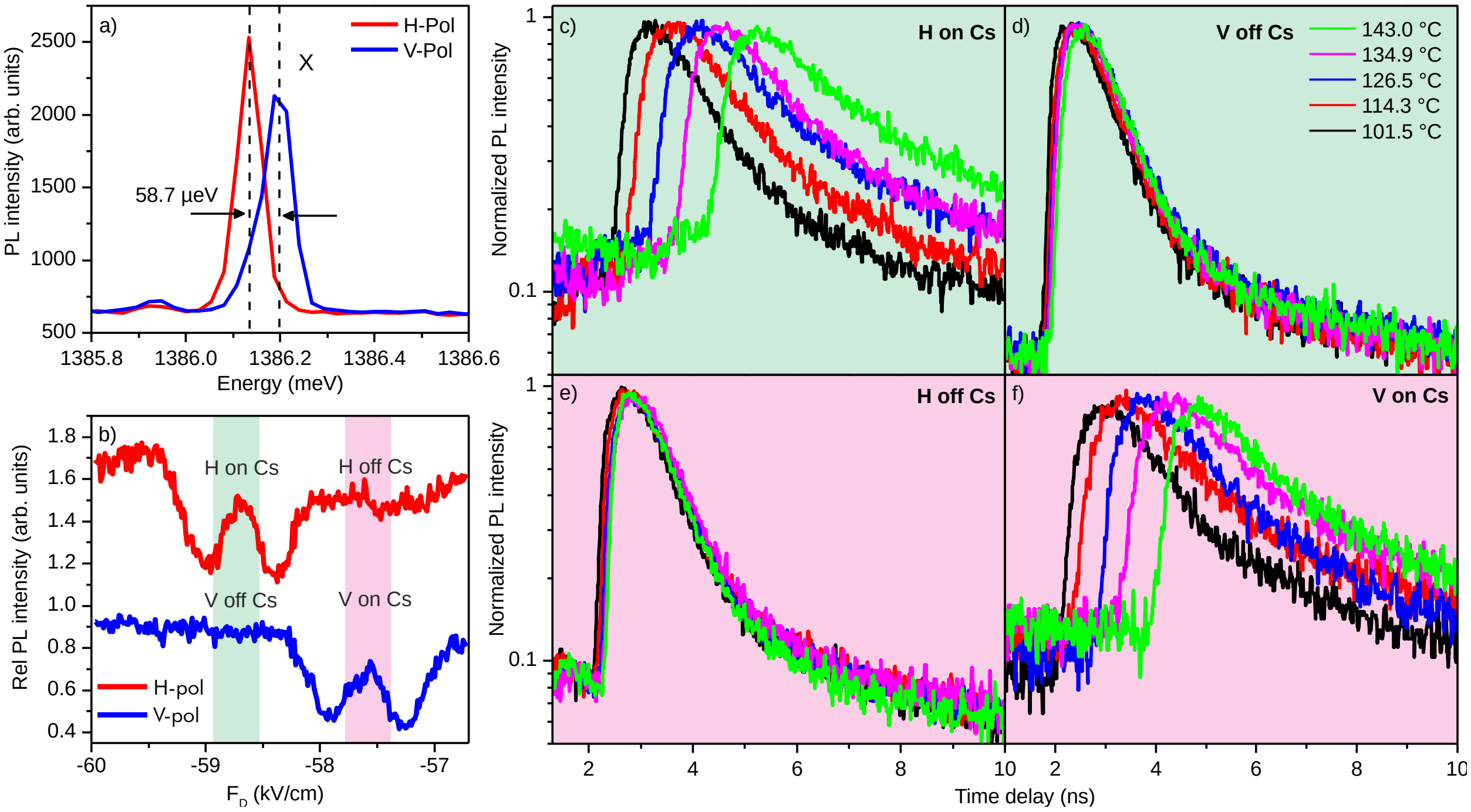

**Figure 3.** (a) Polarization-resolved PL spectra of the neutral exciton featuring a fine-structure-splitting of ~59 µeV. (b) Amplitude of both exciton polarization components tuned through the Cs $D_1$ absorption line via the electric field across the diode. The dips in transmission mark the absorption of the hyperfine structure in Cs. (c) – (f) Time-resolved PL measurements on exciton emission for polarization components of the exciton in resonance with the Cs $D_1$ absorption ((c) and (f)) or out of resonance ((d) and (e)).

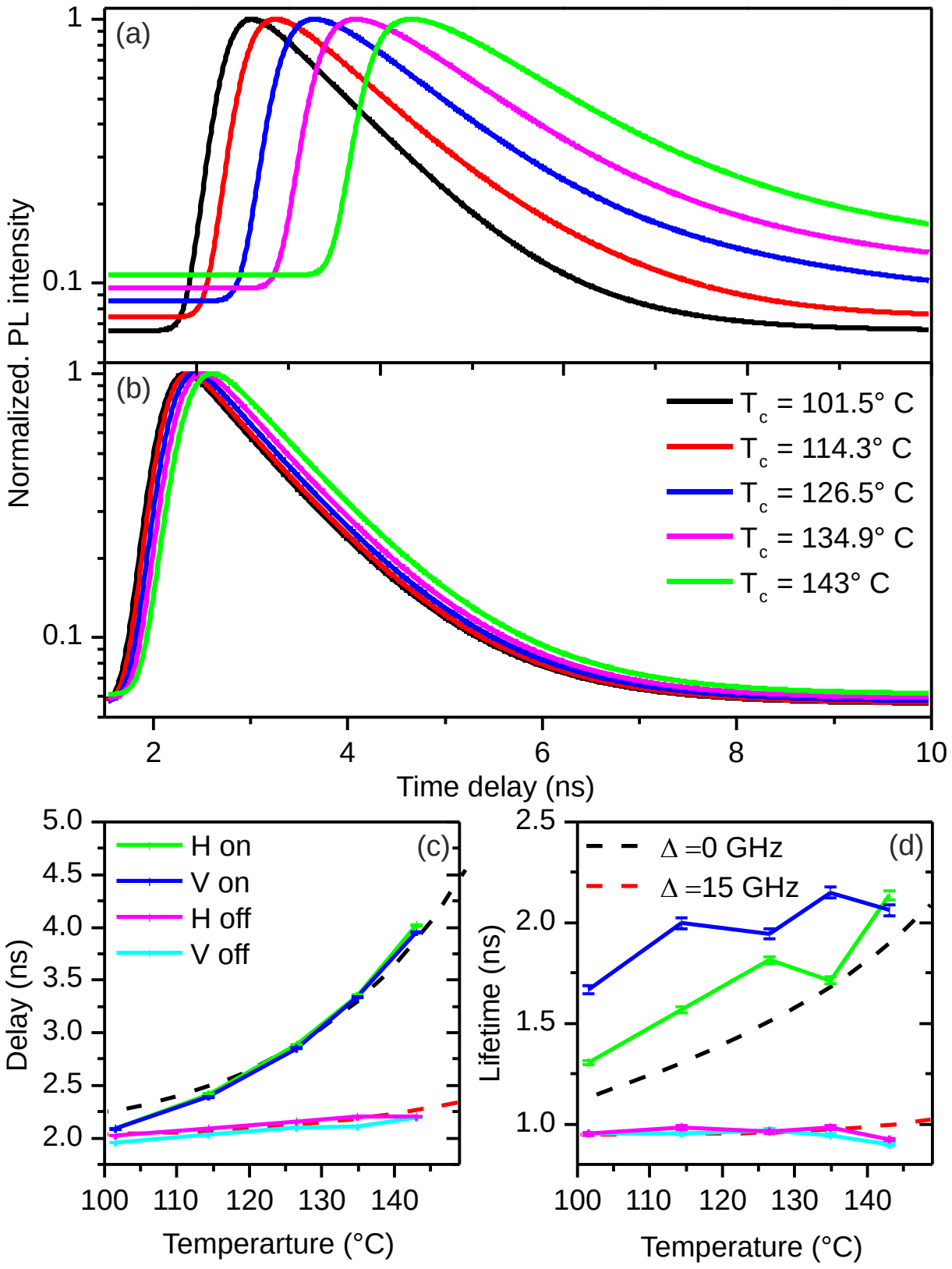

**Figure 4.** (a) and (b) Numerical simulation of a Gaussian-broadened QD emission line with a detuning of 0 (a) and 14 GHz (b) from Cs $D_1$ at cell temperatures between 90°C and 150°C. (c) Temporal delay as extracted from the simulations (dashed lines) and the experimental data (solid lines) and as function of cell temperature. The green, blue and black curves correspond to $\Delta = 0$ GHz while the pink, light blue and red curves show the delay for $\Delta = 14$ GHz. (d) Lifetime vs. cell temperature as extracted from the experiment (solid lines) and the simulations (dashed lines). Please notice the delays obtained from the experiment are shifted to compensate for different optical path lengths. As in (c) green, blue, black and pink, light blue and red lines correspond to 0 and 58 μeV detuning, respectively.

AUTHOR INFORMATION

**Corresponding Author**

* E-mail: johannes.wildmann@jku.at

** E-mail: rinaldo.trotta@jku.at

**Author Contributions**

R.T. and J.S.W. and A. R. conceived and designed the experiment. R.T., J.M.S. and J.S.W. fabricated the devices. J.S.W performed the measurement and data analysis with help from R. T. and A. R. M. O'S., E.Z. and O.G.S. carried out the sample growth. J.S.W. wrote the manuscript with help from all the authors.

**Notes**

The authors declare no competing financial interest.

ACKNOWLEDGMENT

We thank A. Predojević, T. Fromherz, M. Reindl, G. Katsaros and T. Lettner for the fruitful discussions and F. Binder, A. Halliovic, U. Kainz, E. Vorhauer, and S. Bräuer for technical assistance. The work was supported financially by the European Union Seventh Framework Program 209 (FP7/2007-2013) under Grant Agreement No. 601126 210 (HANAS)

REFERIENCES

Supplemental Material for

# "Atomic Clouds as Spectrally-Selective and Tunable Delay Lines for Single Photons from Quantum Dots"

*Johannes S. Wildmann*[†,]*, *Rinaldo Trotta*[†,]**, *Javier Martín-Sánchez*[†], *Eugenio Zallo*[‡], *Mark O' Steen*[§], *Oliver G. Schmidt*[‡], *Armando Rastelli*[†]

[†]Institute of Semiconductor and Solid State Physics, Johannes Kepler University, Altenbergerstr. 69, A-4040 Linz, Austria

[‡]Institute for Integrative Nanosciences, IFW Dresden, Helmholtzstr. 20, D-01069 Dresden, Germany

[§]Veeco Instruments, 4875 Constellation Drive, 55127 St. Paul, Minnesota, USA

**High-resolution spectroscopy**

In order to resolve all 4 $D_1$ transitions of the Cs vapor (see Fig. 1c of the main text) we used the ~15-nm broad pulses of a mode-locked Ti:Sa laser. Before entering the Cs cell, the laser light was spectrally filtered using a volume Bragg grating (VBG) with a 0.2-nm-wide transmission window and a Fabry-Perot interferometer (FPI) with a free-spectral range (FSR) of 41.4 μeV (10 GHz). This combination allows for spectral filtering with a resolution of 0.28 μeV (67 MHz) . The FPI resonance mode (i.e. laser light energy) was then tuned in the spectral range of Cs $D_1$ with a step resolution of 0.083 μeV. After propagation through the Cs cell the laser light was then dispersed with a spectrometer (spectral resolution of ~25 μeV) and detected by a liquid-nitrogen-cooled charge coupled device (CCD)

In order to deconvolve the sharp FPI peaks (which are not resolved by the spectrometer) lorentzian fits were used and their positions and amplitudes (intensities) were followed while scanning the FPI over the whole free spectral range. This procedure leads to a spectral resolution that is 250 times larger than what can be achieved using a single spectrometer. The procedure described above was repeated for different temperatures of the Cs cell, resulting in the spectra reported in Fig. 1c.

**Numerical Simulations of the group velocity and temporal distribution of the transmitted photons.**

In order to simulate the absorption spectrum of Cs we calculate the energy shift of the hyperfine splitting in Cs $D_1$ depending on the total atomic angular momentum **F** and the total nuclear angular momentum **I**. A Gaussian distribution with a Doppler broadened linewidth was used to simulate each of the resonances and collisional broadening due to vapor pressure and temperature was taken into account[i,ii].

The susceptibility for Cs $D_1$ is given by the following expression

$$\chi(\nu)=A\left(\frac{g_{33}}{\nu_{33}-\nu-i\,\gamma}+\frac{g_{34}}{\nu_{34}-\nu-i\,\gamma}+\frac{g_{43}}{\nu_{43}-\nu-i\,\gamma}+\frac{g_{44}}{\nu_{44}-\nu-i\,\gamma}\right)$$

where $g_{33}$,$g_{34}$,$g_{43}$,$g_{44}$ are the oscillator strengths for each transition and $\nu_{33}$, $\nu_{34}$,$\nu_{43}$,$\nu_{44}$ are the resonance frequencies [iii]. The parameters A (amplitude), and $\gamma$ (damping parameter) are left as fitting parameters (see the following).

The refractive index can be expressed as follows

$$n(\nu)=\sqrt{1+\chi(\nu)}=n_r(\nu)+i\kappa(\nu)$$

The fitting parameters A and $\gamma$ at each specific cell temperature are obtained after fitting the simulated [ii] absorption coefficient κ, which in turn provides the real part of the refractive index $n_r$.

The group velocity of Cs $D_1$ can be then obtained as a function of the frequency using the following formula[iv]

$$v_g(\nu)=\frac{1}{n(\nu)+\nu\frac{d\,n(\nu)}{d\,\nu}}$$

Besides the spectral shape of the group velocity, we are also interested in the temporal distribution of QD photons before and after propagation trough the Cs cell. The initial temporal distribution of the QD PL can be modeled using the following formula[v]

$$A(\tau)=y_0+\frac{B}{t_0}e^{\left(\frac{1}{2}\left(\frac{\omega_t}{t_0}\right)^2-\frac{x-x_c}{t_0}\right)}\left(\frac{1}{2}+\frac{1}{2}\mathrm{Erf}\left(\frac{x-x_c}{\omega_t}-\frac{\omega_t}{t_0}\right)\right)$$

where $\mathrm{Erf}\left(\frac{x-x_c}{\omega_t}-\frac{\omega_t}{t_0}\right)$ is the Gaussian error function, B is the amplitude, $y_0$ an offset in the intensity, $t_0$ is a measure for the temporal decay, $\omega_t$ is the temporal width and $x_c$ its position in time. The parameters for this function were found by fitting the experimental data at 101.7 °C and for large detuning $\Delta$=14.2 GHz.

The final temporal distribution of the photons at a specific temperature of the Cs cell is calculated in a two-step process. First, the group velocity and the optical transmission are calculated for given Δ. Second, the QD PL temporal distribution is obtained by integrating over all Δ and by weighting the results using a Gaussian spectral distribution for the QD with linewidth ω obtained from the experimental data.

**Absorption properties vs. QD linewidth, cell temperature and detuning.**

The PL intensity transmitted through the Cs cell as a function of time is simulated for given QD linewidth, cell temperature and detuning (Fig. S1). Fig. S1 a – d show the simulated time-resolved decay curves for temperatures ranging from 50°C to 150 °C, and a QD linewidth ω of 2 GHz (a), 5 GHz (b), 10 GHz (c) and 15 GHz (d). The detuning of the QD emission with respect to the Cs vapor was set to 0 GHz for all the simulations except the one shown in the inset of Fig. S1a, which corresponds to a detuning of 15 GHz. For $\Delta = 0$, the data clearly shows that the peak position of the decay curves, i.e., the average time delay, does not depend on the spectral broadening of the QD transitions. However, the temporal broadening as well as the optical absorption increases with the spectral linewidth. Finally, it is interesting to note that the substantial temporal delay of 0.37 ns with negligible optical absorption can be observed when the QD line is tuned out of the center of the D1 lines, as displayed in the inset of Fig S1 (a) for Δ=15 GHz. Despite this can be qualitatively explained taking into account the behavior of the group velocity shown in the Fig.1 of the main text, it is interesting to investigate how the optical absorption changes as a function of the delay for different spectral linewidth, and cell temperatures. This is shown in Fig. S1 (e) for $\Delta = 0$, where we observe that the drop in transmission with the delay changes from almost linear for small spectral linewidths to a superposition of an exponential and a linear decay for large spectral linewidths. In the ideal case of Fourier limited photons featuring a line width $\omega \approx 0.5\ GHz$ the absorption is less than 1% while the maximal temporal delay is still of the order of 2.5 ns. For different $\Delta$ (see Fig. S1 (f)-(g)), a similar behavior is observed although in the case of $\Delta = 15$ GHz significant delay can be achieved only for very high temperature of Cs vapor.

On the one hand, the detailed analysis reported here can be used to explain all the features observed in the experiment, as for example the appearance of un-delayed peaks, pronounced absorption, and temporal broadening clearly visible for large spectral linewidth (see Fig. 1(e) and [vi]). One the other hand, this analysis also suggests that for broad linewidth it is possible to observe a small temporal delay with negligible absorption and broadening by tuning the QD far for the hyperfine doublet.

**Extracting QD linewidth through QD-Cs transmission**

The linewidth of the QD optical transitions was extracted from the data reported in Fig. 2b, that result from scanning a QD with a given spectral linewidth $\omega$ through the hyperfine structure of Cs. To extract the QD linewidth, we perform the following steps: ($i$) the reference transmission spectrum of Cs was measured by high-resolution spectroscopy (as discussed above) at the temperature of 136 °C (temperature used for the measurements reported in Fig. 2b). ($ii$) We convolved the result obtained in ($i$) with a QD transition featuring a Gaussian lineshape and a linewidth $\omega$. (*iii*) We fit the experimental data by changing $\omega$ and by performing the least squares minimization.

**Sample and device fabrication**

The sample was grown by Molecular Beam Epitaxy (MBE) using a commercial Omicron 4-inch MBE machine. A 200 nm-thick undoped GaAs buffer layer was grown at a substrate temperature $T_{sub}$=590 °C on semi-insulating GaAs (001) followed by a 100 nm-thick $Al_{0.75}Ga_{0.25}As$ sacrificial layer. Afterwards, a tri-layer diode structure [180 nm-thick n-doped GaAs layer/150 nm-thick GaAs intrinsic layer containing InAs QDs/100 nm-thick p-doped GaAs layer] was grown. The InGaAs QDs were grown at 500°C and capped by an indium flush technique[vii]. Selected areas on the wafer with a QDs density as low as $10^7$ $cm^{-2}$ were used for further processing. The GaAs nanomembranes containing QDs were obtained by optical lithography and selective chemical etching and integrated onto a 300-µm-thick $[Pb(Mg_{1/3}Nb_{2/3})O_3]_{0.72}$ – $[PbTiO_3]_{0.28}$ (PMN-PT) piezoelectric substrate by gold thermocompression bonding. The electric field across the diode structure (across the piezo actuator) was applied to control the QD energy levels via the quantum confined stark effect (strain). Further details of the processing and device fabrication can be found elsewhere [viii].

**Measurement setup**

The optical experiments were performed using a micro luminescence setup equipped with a 50x objective (NA=0.42) for optical excitation and PL collection. For excitation we used a pulsed TiSa-Laser with a repetition rate of 80MHz and a typical pulse width of the order of 100fs. The Laser is tunable in the range between 690 and 1080 nm, which allows for optical excitation in the QDs either above band-gap or in the wetting-layer. For low temperature measurements (8K), the sample was mounted in a He flow cryostat. The Cs vapor cell was inserted in the PL optical path and consists of a 7.5 cm long glass cell filled with Cs vapor and wounded with heating foils to set precisely the temperature (temperature accuracy ~ 0.1 °C). The PL spectra are measured by a liquid-nitrogen-cooled Si CCD detector. Single photon avalanche photodiodes were used for time-resolved PL measurements. The polarization properties of the emitted light are analyzed using a rotatable lambda half wave plate combined with a fixed linear polarizer.

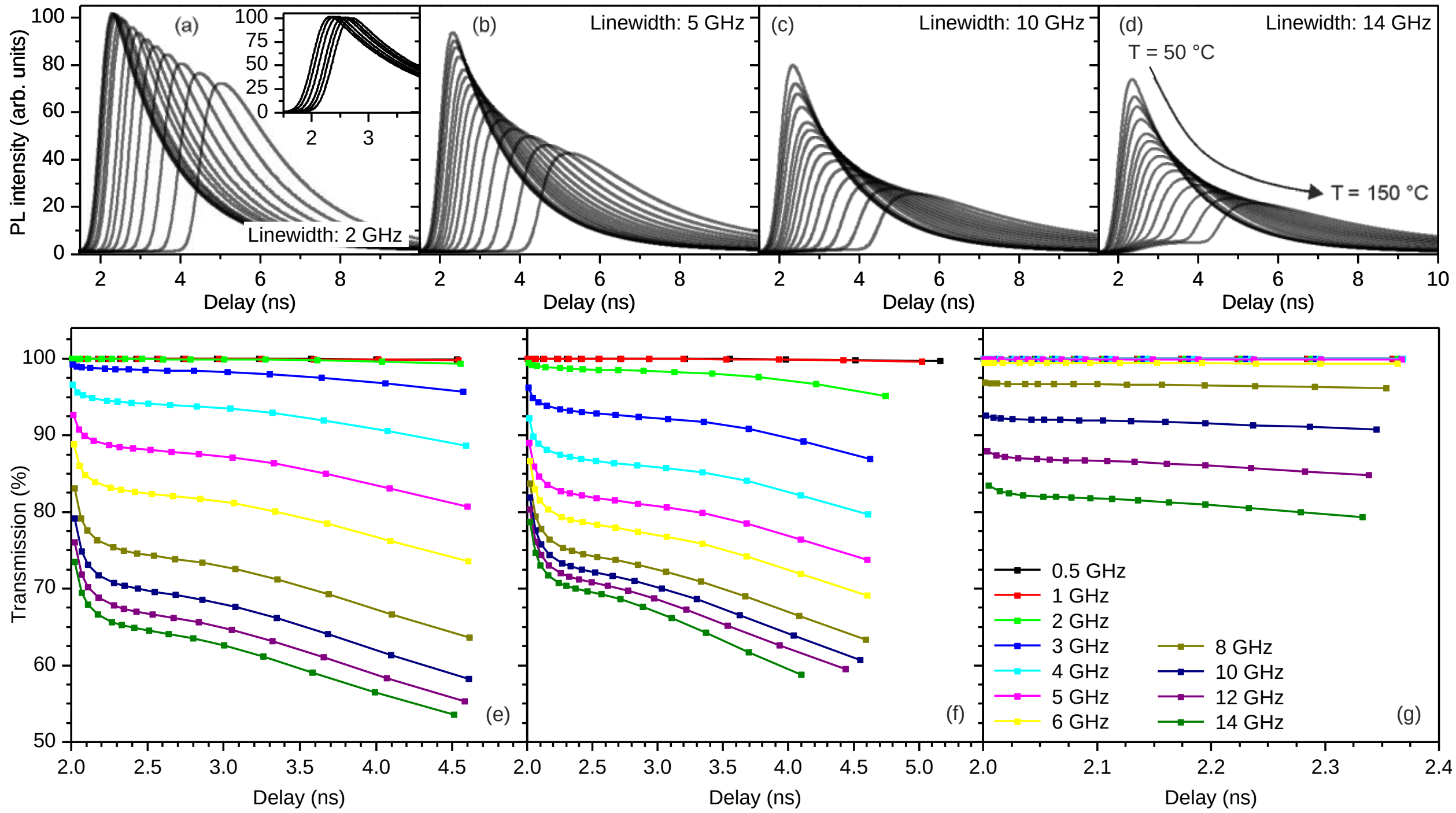

**Figure S1.** (a) - (e ) Simulated time-resolved PL decay curves as a function of the temporal delay, for temperatures ranging from 50 °C to 150 °C, and photon linewidth ω values of 2 GHz (a), 5 GHz (b), 10 GHz (c) and 14 GHz (d). (a,b,c,d) and (a inset) present the decay curves for $\Delta$=0 GHz and 15 GHz, respectively. (e), (f) and (g) display the optical transmission as a function of the temporal delay for different $\omega$(ranging from 0.5 $GHz$ to 14 GHz) and $\Delta$=0, 2, 15 GHz.

AUTHOR INFORMATION

**Corresponding Authors**

*E-mail: johannes.wildmann@jku.at

** E-mail: rinaldo.trotta@jku.at

ASSOCIATED CONTENT

(Word Style "TE_Supporting_Information"). **Supporting Information**. This material is available free of charge via the Internet at http://pubs.acs.org